\documentclass[twocolumn,showpacs,aps,prb,amsmath,amssymb]{revtex4}
\usepackage{graphicx}% Include figure files
\begin{document}

\title{Excitations of the bimodal Ising spin glass on the brickwork lattice}

\author{W. Atisattapong}
\affiliation{Department of Mathematics, Faculty of Science,
Mahidol University, Rama 6 Road, Bangkok 10400, Thailand}

\author{J. Poulter}
\affiliation{Department of Mathematics, Faculty of Science,
Mahidol University, Rama 6 Road, Bangkok 10400, Thailand}

\date{\today}

\begin{abstract}
An exact algorithm is used to investigate the distributions of the
degeneracies of low-energy excited states for the bimodal Ising spin glass
on the brickwork lattice. Since the distributions are extreme and do not
self-average, we base our conclusions on the most likely values of the
degeneracies. Our main result is that the degeneracy of the first excited
state per ground state and per spin is finite in the thermodynamic limit.
This is very different from the same model on a square lattice where a
divergence proportional to the linear lattice size is expected.
The energy gap for the brickwork lattice is obviously $2J$ on finite systems
and predicted to be the same in the thermodynamic limit. Our results
suggest that a $2J$ gap is universal for planar bimodal Ising spin glasses.
The distribution of the second contribution to the internal energy has a mode close to zero and we
predict that the low-temperature specific heat is dominated by the leading
term proportional to $T^{-2} \exp(-2J/kT)$.

\end{abstract}


\maketitle
\section{introduction}
Although bimodal planar Ising spin glass models are very simple in concept,
they are extremely controversial. One main reason why concerns the energy gap
between the ground and first excited states. To date, most work has been
reported for the square lattice where there is wide acceptance for the
scenario of a critical point only at zero temperature \cite {HY01,H01}.
In the absence of any contradictory evidence or suggestion, we assume this
to be the case for other planar models, including the brickwork lattice.

Bimodal models have bond (nearest-neighbor) interactions of fixed magnitude
$J$ and random sign. Both the ground and excited states have a very large
degeneracy. For an infinite square lattice, without open
boundaries, it is clear that any finite number of spin flips must either
result in another ground state or an excited state with an increase in
energy not less than $4J$. For the brickwork lattice the corresponding
energy gap is clearly $2J$, since the lattice coordination number is three.

About 20 years ago, Wang and Swendsen \cite {WS88} published evidence
that the energy gap for the square lattice in the thermodynamic limit was
$2J$. This flew in the face of the naive expectation of $4J$. Essentially,
the claim was that it is possible for an infinite number of spin flips to
provide an excited state with energy only $2J$ above a ground state. The
issue here is the noncommutativity of the zero-temperature and thermodynamic
limits. The thermodynamic limit has to be taken first. Nevertheless, for the
brickwork lattice we do not expect this to be an issue and the main
interest of this paper is to show evidence that this is the case. Both models
have the same energy gap in the thermodynamic limit, although the reasons
why are quite different.

For the square lattice there are three scenarios in the literature regarding
the energy gap. First, support for the $2J$ energy gap includes work at
finite temperatures involving exact computations of partition functions
\cite {LGM04}, a worm algorithm \cite {W05} and Monte Carlo simulation
\cite {KLC05}. Distributions of excited-state degeneracies at arbitrary
temperature \cite {Wanyok} also indicate a $2J$ gap. Essentially, it was
shown that the degeneracy of the first excited state per ground state and
per spin diverges in the thermodynamic limit. In consequence the $4J$ gap
of the finite system is reduced to $2J$. 

Second, Saul and Kardar \cite {SK93} reported that the energy gap should be
$4J$ as suggested by simple analysis. The third published scenario
\cite {JLM06} basically claims that the energy gap approaches zero in the
thermodynamic limit leading to power law behaviour for the specific heat.
This possibility has been discussed at some length in Ref. \onlinecite {KLC07},
although clear conclusions remain unavailable due to difficulties related
to finite lattice size and extrapolation to very low temperature.

\begin{figure}[h]
\includegraphics*[width=8.5cm]{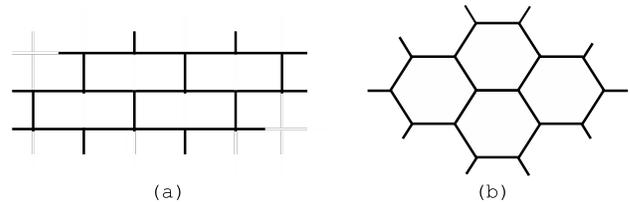}
\caption{\label{f:Fig1}(a) The brickwork lattice. (b) The equivalent hexagonal lattice.}
\end{figure}

The brickwork lattice (Fig. \ref{f:Fig1}(a)) studied in this paper is logically equivalent to the
hexagonal, or honeycomb, lattice (Fig. \ref{f:Fig1}(b)). Very little work has been published to date,
especially concerning the ground state. The ground state energy per bond
has been quoted \cite {VL97} as $-0.82J$. The entropy per spin has been
reported by Aromsawa \cite {Anuchit} as $0.02827(5)k$ with an energy
$-1.2403(2)J$ per spin, in good agreement with Ref. \onlinecite {VL97}.
The spin glass phase is thought to exist for a concentration of negative
bonds \cite {Bendisch} above about $0.067$. Work at finite temperature
\cite {deQ06,NO06,Ohzeki} places the multicritical, or Nishimori, point
at the same concentration; in agreement if reentrant phase boundaries are
absent.

Our calculations of the degeneracies of excited states for the brickwork
lattice are exact. The temperature is fixed and arbitrarily low; we do not
use any numerical value. The lattice is constructed by taking a square
lattice and diluting bonds in a regular manner to leave plaquettes with six
perimeter bonds; logically equivalent to hexagons as shown in Fig. \ref{f:Fig1}. The disorder realizations
are independently quenched random configurations of negative bonds in a
patch that contains all the frustrated plaquettes. Periodic boundary
conditions are used in one dimension. The number of sites $L$ for this
dimension is necessarily even. The cylindrically wound frustrated patch is
embedded in an infinite unfrustrated environment in the second dimension.
%END JULIAN
\section{formalism}
\begin{figure}[h]
\includegraphics*[width=7cm]{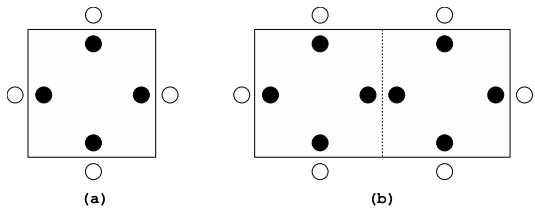}
\caption{\label{f:Fig2}(a) A square plaquette as in Ref. \onlinecite {PB05}. (b) A brickwork plaquette obtained by dilution of the central bond (denoted by a dotted line) between two square plaquettes.}
\end{figure}

An algorithm based on the Pfaffian method \cite {GH64} and degenerate state perturbation theory \cite {B82,BP91,PB05} for the square lattice was adapted to evaluate the degeneracies of excited states for the brickwork lattice. The main points of this procedure are given in the following. From the square lattice, we dilute bonds in a regular manner to define the brickwork lattice. Using the fermion decoration of bonds (one either side), a brickwork plaquette has eight fermions inside (filled circles) and six others across the bonds as shown in Fig. \ref{f:Fig2}. 

As for the bimodal Ising spin glass on the square lattice, the partition function is given by $Z \sim (\det D)^{1/2}$ where $D$ is a real skew-symmetric $4N \times 4N$ matrix for a lattice with $N$ sites. $D$ is a singular matrix at zero temperature with zero eigenvalues which are equal in number to the number of frustrated plaquettes. Each eigenvalue approaches zero according to the form
\begin{equation}
\epsilon=\pm\frac{1}{3}X\exp(-2Jr/kT),
\label{e:EV}
\end{equation}
where $r$ is an integer (an order of perturbation theory) and $X$ is a real number. The quantities $r$ and $X$ can be exactly evaluated by degenerate state pertubation theory. The ground-state energy and entropy of the system can be defined similarly as for the square lattice \cite{BP91}. It is equivalent to expressing the ground-state degeneracy as
\begin{equation}
M_0=\prod X.
\label{e:M}
\end{equation}
where the product is over all the positive defect eigenvalues.

\begin{figure}[h]
\includegraphics*[width=4cm]{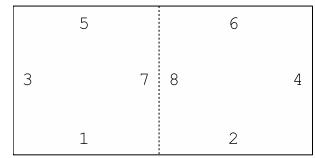}
\caption{\label{f:Fig3}The states in the bond basis of a brickwork plaquette. It is convenient to use the bond basis introduced in Ref. \onlinecite{Wanyok}, that is $\mid\pm\rangle=\frac{1}{\sqrt{2}}(\mid\alpha\rangle\pm\zeta\mid\beta\rangle)$ where $\mid\alpha\rangle$ and $\mid\beta\rangle$ are shown in Fig. 4 of Ref. \onlinecite{BP91} and $\zeta$ represents the sign of the bond.}
\end{figure}

The gauge-invariant method is applied similarly as for the square lattice \cite{BP91} to separate the singularities of $D$ for the brickwork lattice using real matrices as follows. At zero temperature, the perfect system (no frustration) Green's function \cite{BP91} $g_0$ is obtained by transforming $D$ into a plane-wave basis and inverting an $8\times8$ matrix. The nonzero elements of $g_0$ are only across bonds and localized inside plaquettes. Across bonds we have $g_0(+,-)=-g_0(-,+)=\frac{1}{2}$ where the matrix indices are defined in Fig. 4 of Ref. \onlinecite{BP91}. Within a plaquette $g_0$ is as given by the following matrix \cite{Anuchit}:
\begin{equation}
g_{0}=\frac{1}{2}\left[
\begin{array}{cccccccc}
 0& -1& 1&-1&1&1 &-2&0\\
 1& 0& 1& -1&1&1&0&2\\
 -1&-1& 0&-1& 1&1&-2&0\\
1& 1& 1& 0&1&1&0&2\\
-1&-1& -1&-1& 0& 1&-2&0\\
-1&-1& -1&-1&-1& 0&0&-2\\
2&0& 2&0& 2& 0&0&2\\
0&-2&0& -2& 0& 2&-2&0
\end{array}
\right]
\label{e:g0}
\end{equation}
where the elements of $g_0$ are with respect to the bond basis of a plaquette as shown in Fig. \ref{f:Fig3}.

\begin{figure}[h]
\includegraphics*[width=5.3cm]{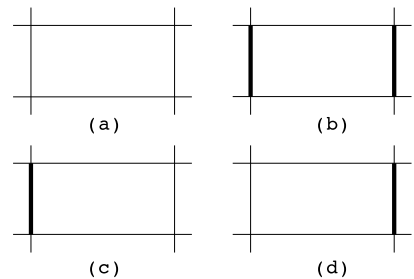}
\caption{\label{f:Fig4}Heavy lines denote negative (defect) bonds. There are four types of possible plaquettes after gauge transformation to vertical defect bonds. (a) and (b) are unfrustrated plaquettes while (c) and (d) are frustrated plaquettes.}
\end{figure}

We set the bimodal problem for the brickwork lattice by placing negative bonds at random. To reduce the complexity, we perform gauge transformations as in Ref. \onlinecite{BP91} so that the negative (defect) bonds are vertical only. We classify plaquettes into four types as shown in Fig. \ref{f:Fig4}.

To determine ground state properties, degenerate state perturbation theory is applied at arbitrarily small finite temperatures. We write exactly $D=D_{0}+\delta D_{1}$ where $\delta=1-t$, with $t=\tanh(J/kT)$, is used as a parameter for a perturbation expansion. The matrix $D_{0}$ is singular and has defect eigenvectors $\mid d \rangle$ corresponding to defect (zero) eigenvalue, that is $D_{0}\mid d\rangle=0$, localized on each frustrated plaquette, similar to those found for the square lattice \cite{BP91}. 

At first order the matrix $D_{1}$, which is $2\times 2$ block diagonal across bonds, is diagonalized in the basis of the defect eigenvectors. We also require the continuum Green's function $G_{c}=(1-P)g_c(1-P)$ where $P=\sum_d\mid d\rangle\langle d\mid$ and $g_c=g_0+g_0Ug_0$ with $U$ defined similarly to Ref. \onlinecite{BP91}. We can write $g_c=g_{c1}+g_{c2}$ where $g_{c1}$ has matrix elements connecting the basis states within a brickwork plaquette while $g_{c2}$ connects only basis states across a single bond, that is $g_{c2}(+,-)=-g_{c2}(-,+)=\frac{1}{2}$ and is only relevant for excited states. Although $g_0$ can take us to the fermions associated with diluted bonds (labeled $7-8$ in Fig. \ref{f:Fig3}), it can be proven that these matrix elements do not effect the calculation of the partition function since $D_1$ across that bond is equal to zero (there is no energy); we can disregard them. An alternative Pffaffian based on three nodes per site has been given in Ref. \onlinecite{GH64} but it cannot be adapted to our defect problem.

The matrix $g_{c1}$ can be considered in the subbasis with only six fermions (labeled $1-6$ in Fig. \ref{f:Fig3}) without any change of the gauge-invariant ground or excited state properties. We can also arrange to have $g_{c1}$ orthogonal to defect states, that is $g_{c1}\mid d \rangle=0$, by understanding that we can add any matrix $A$ to $g_{c1}$ as long as $(1-P)A(1-P)=0$. This reduces the number of arithmetic operations for the calculation of excitations. The matrix $g_{c1}$ can be presented for an unfrustrated plaquette as
\begin{equation}
g_{c1}=\frac{1}{2}\left[
\begin{array}{cccccc}
 0& -1& 1&-1&s&s\\
 1& 0& 1& -1&s&s\\
 -1&-1& 0&-1& s&s\\
1& 1& 1& 0&s& s\\
-s&-s& -s&-s& 0& 1\\
-s&-s&-s& -s& -1& 0
\end{array}
\right]
\label{e:gc1u}
\end{equation}
where $s=1$ for an unfrustrated plaquette with no negative bond (Fig. \ref{f:Fig4}(a)) and $s=-1$ for an unfrustrated plaquette with two negative bonds (Fig. \ref{f:Fig4}(b)). The matrix $U$ only occurs for plaquettes with two defect bonds. In detail, $U_{34}=-U_{43}=-2$. The matrix $g_{c1}$ for a frustrated plaquette is given by
\begin{equation}
g_{c1}=\frac{1}{6}\left[
\begin{array}{cccccc}
  0& -2&2& -1 & -s&  0\\
2&  0& 1&-2&  0& s\\
 -2&  -1&  0&  0&-2s& -s\\
 1&2&  0&  0& s&2s\\ 
 s&  0&2s&  -s&  0&2\\
  0& -s& s& -2s& -2&  0
\end{array}
\right]
\label{e:gc1f}
\end{equation}
where $s=1$ for a frustrated plaquette with a left negative bond (Fig. \ref{f:Fig4}(c)) and $s=-1$ otherwise (Fig. \ref{f:Fig4}(d)). The corresponding defect states are
\begin{equation}
\mid d \rangle=\frac{1}{\sqrt6}\left(\mid1\rangle+\mid2\rangle+\mid3\rangle+\mid4\rangle-s\left(\mid5\rangle+\mid6\rangle\right)\right)
\end{equation}
The prefactor in Eq. (\ref{e:EV}) is essentially determined by the normalization of this vector.

At second order, we diagonalize $D_{2}=D_{1}g_{c1}D_{1}$. To continue for higher orders, we require the Green's functions $G_r$, as given in Ref. \onlinecite{Wanyok}, obtained from previous orders; that is for states whose degeneracy has already been lifted. The general rule for $D_r$ (at $r$th order) can be expressed as $D_{r}=D_{r-1}(1+G_{r-2}D_{r-2})\ldots(1+G_{1}D_{1})g_{c1}D_{1}$ . 

The calculation of the internal energy and the specific heat for the brickwork lattice is similar to the square lattice as described in Ref. \onlinecite{Wanyok}. The internal energy is given by
\begin{equation}
U=\sum_{m=0}^{\infty}e^{-2Jm/kT}U_m.
\label{e:U}
\end{equation}
where $U_0$ is the ground state energy and $U_m=-2^{m}J\mathrm{Tr}R^m$, for $m>0$, and 
\begin{equation}
R=D_1g_{c1}\left(1+D_1G_1\right)\left(1+D_2G_2\right)\ldots\left(1+D_{r_{max}}G_{r_{max}}\right).
\label{e:R}
\end{equation}
$r_{max}$ is the highest order of pertubation theory required.

However, there is one essential distinction between the square and brickwork lattices. Since we need three colors to color the brickwork lattice, this means that the color rules described in Ref. \onlinecite{Wanyok} are invalid and it follows that $U_m \neq 0$ for all $m$. The specific heat per spin can be expressed in terms of the internal energy as
\begin{equation}
c_v=\frac{1}{N}\frac{dU}{dT}=\frac{1}{N}\left(\frac{2J}{kT^2}\right)\sum_{m=1}^{\infty}me^{-2Jm/kT}U_m.
\label{e:SP}
\end{equation}
The degeneracy of the $i$th excited state is given as $M_i$. Expanding $\ln Z$, we get, for example, $U_1=2J(\frac{M_1}{M_0})$ and $U_2=4J(\frac{M_2}{M_0}-\frac{1}{2}(\frac{M_1}{M_0})^2)$.

\section{results}
\begin{figure}[ht]
\includegraphics*[angle=-90,width=8.5cm]{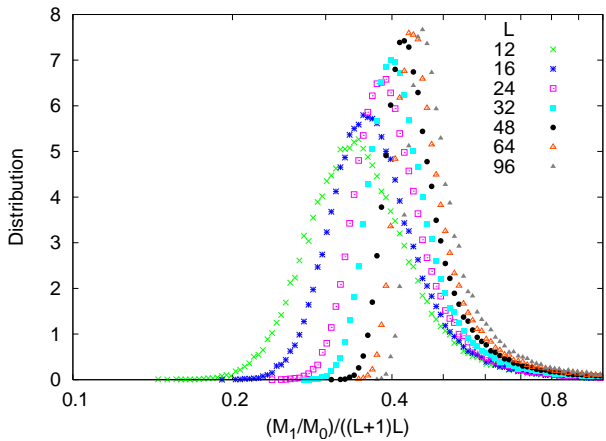}
\caption{\label{f:Fig5}(Color online) Distributions for $\frac{M_1}{M_0}\frac{1}{(L+1)\times L}$ with various values of lattice size $L$ with $(L+1)\times L$ spins. Each distribution includes $10^5$ disorder realizations, except for $L=96$ which has $30,000$.}
\end{figure}
\begin{figure}[ht]
\includegraphics*[angle=-90,width=8.5cm]{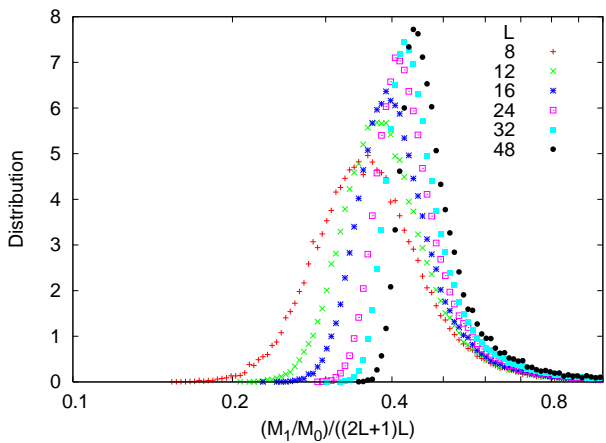}
\caption{\label{f:Fig6}(Color online) Distributions for $\frac{M_1}{M_0}\frac{1}{(2L+1)\times L}$ with various values of lattice size $L$ with $(2L+1)\times L$ spins. Each distribution includes $10^5$ disorder realizations.}
\end{figure}
Fig. \ref{f:Fig5} shows the distributions of $\frac{M_1}{M_0}$ for the spin glass with $(L+1)\times L$ spins. It is clear that the most likely value scales as the number of spins. We also consider a different shape of the patch boundary by changing the lattice size to $(2L+1)\times L$ so that the equivalent hexagonal lattice has a more balanced shape. The distribution of $\frac{M_1}{M_0}$ with $(2L+1)\times L$ spins also shows again a most likely value that scales as the number of spins as shown in Fig. \ref{f:Fig6}. Also, the value is roughly the same. In both cases, it is clear that the leading term of the specific heat grows like $c_v\sim T^{-2}\exp(-2J/kT)$ indicating a $2J$ excitation gap. Moreover, the distributions of $\frac{M_1}{M_0}$ are extreme, do not self-average and are neither normal or lognormal. This is consistent with the square lattice as in Ref. \onlinecite{Wanyok}.

\begin{figure}[ht] 
\includegraphics*[angle=-90,width=6.5cm]{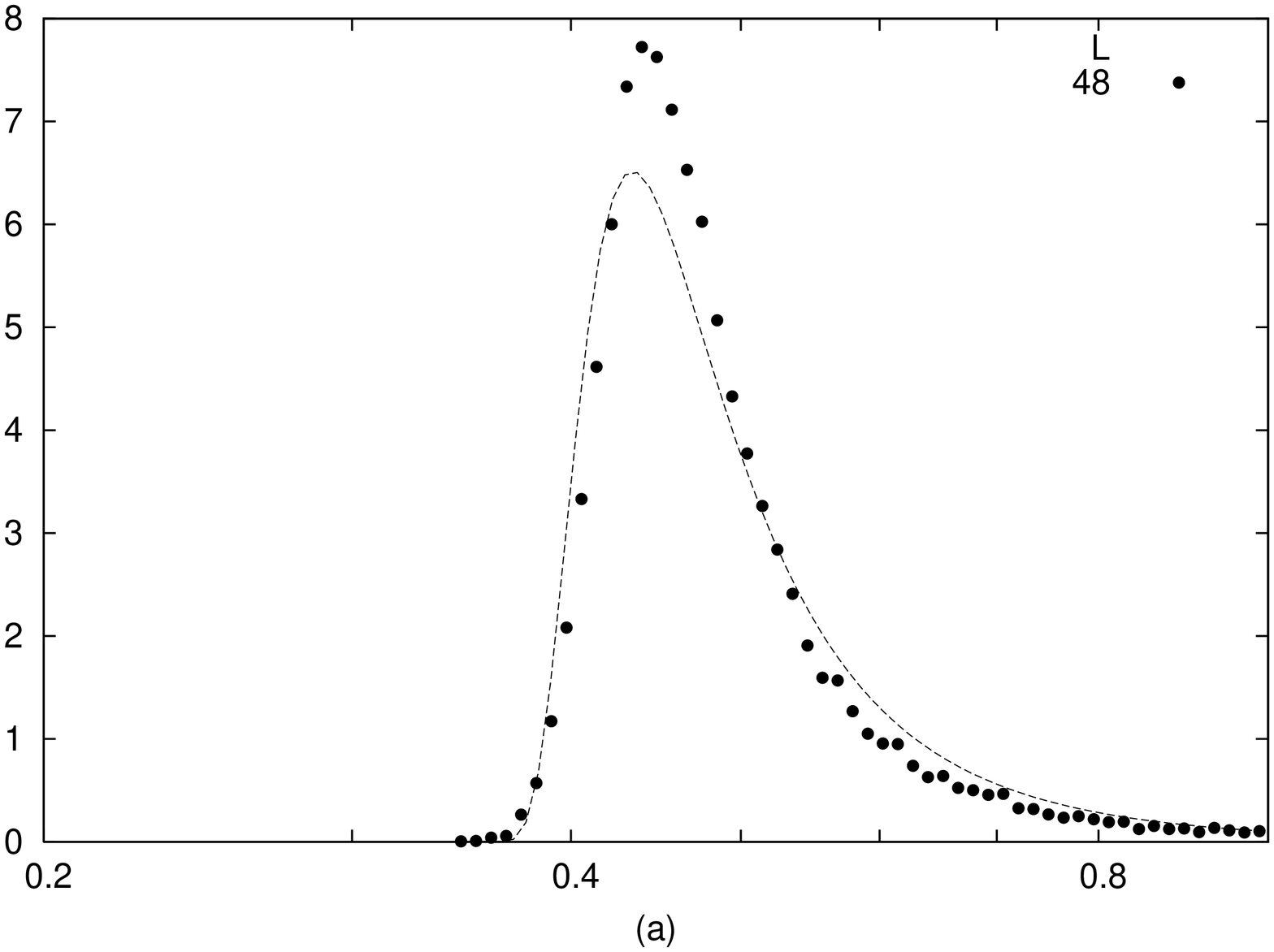}
\includegraphics*[angle=-90,width=6.5cm]{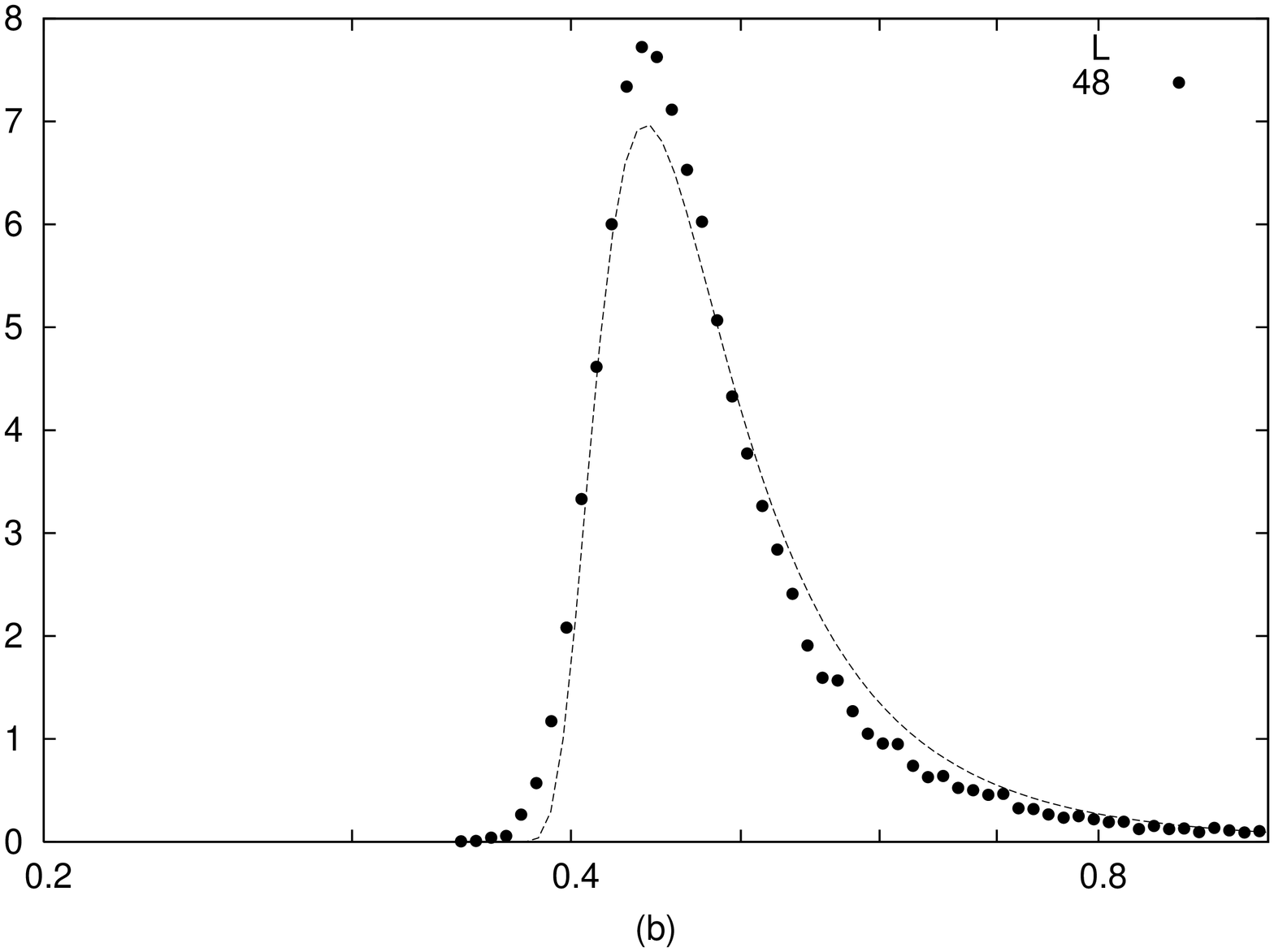}
\caption{\label{f:Fig7}Distributions for $\frac{M_1}{M_0}\frac{1}{(2L+1)\times L}$ with $L=48$ and $(2L+1)\times L$ spins. (a) The line uses the parameters obtained from the algorithm of Hosking \cite{Hosking} fitted to actual disorder realizations. (b) The line uses the parameters obtained from the algorithm of Levenberg-Marquardt \cite{LMmethod} fitted to bin data using the Hosking parameters as initial values.}
\end{figure}

\begin{figure}[ht]
\includegraphics*[angle=-90,width=8.5cm]{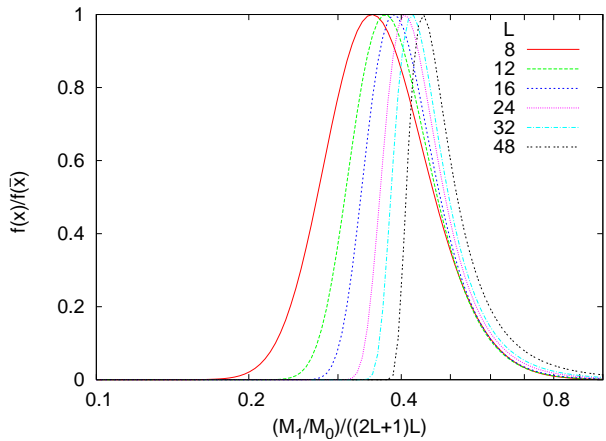}
\caption{\label{f:Fig8}(Color online) Fr\'{e}chet distributions from fitting the bin data in Fig. \ref{f:Fig6} by the algorithm of Levenberg-Marquardt using the Hosking parameters as initial values.}
\end{figure}

\begin{figure}[ht]
\includegraphics*[angle=-90,width=8.5cm]{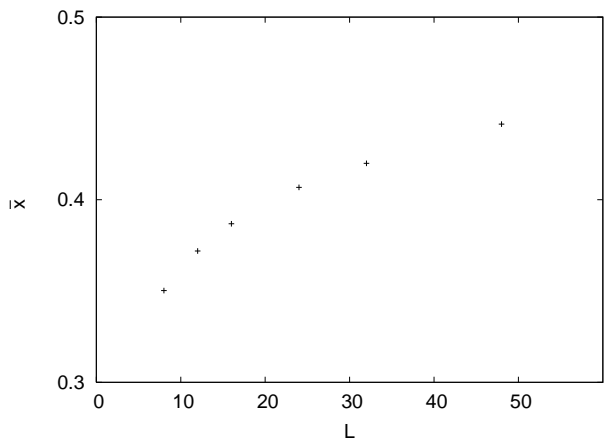}
\caption{\label{f:Fig9}The mode of the fitted distributions with various values of lattice size $L$ with $(2L+1)\times L$ spins.}
\end{figure}

The distributions of $\frac{M_1}{M_0}$ clearly show extreme-value distributed properties with a fat tail. It is thus reasonable to analyze the distributions according to the Fr\'{e}chet distribution with the probability density function:
\begin{equation}
f_{\xi,\mu,\beta}(x)=\frac{1}{\beta}
\left(1+\xi\frac{x-\mu}{\beta}\right)^{-\frac{1}{\xi}-1}
\exp\left(-\left(1+\xi\frac{x-\mu}{\beta}\right)^{-\frac{1}{\xi}}\right)
\label{e:fre}
\end{equation}
where the parameters $\mu$, $\beta$ and $\xi$ indicate location, shape and scale of the distribution respectively. The mode can be calculated by $\overline{x}=\mu+\beta\frac{(1+\xi)^{-\xi}-1}{\xi}$. We estimate the parameters by a maximum likelihood estimator \cite{Hosking} to fit actual disorder realizations. It is found that the Fr\'{e}chet distribution cannot fit exactly the peak of our actual distribution; the quality of the fit is quite poor and also deteriorates with respect to increasing $L$. An example for $L=48$ with $(2L+1)\times L$ spins is shown in Fig. \ref{f:Fig7}(a). We have also tried to polish the fit using the Levenberg-Marquardt method \cite{LMmethod} to fit the bin data, setting initial values of parameters equal to the values obtained from the algorithm of Hosking \cite{Hosking}. This provides alternative parameters and the curve is shown in Fig. \ref{f:Fig7}(b). Although the quality of the fit looks a little better and the position of the mode somewhat improves, we are not convinced that Eq. (\ref{e:fre}) should necessarily describe the distribution exactly. 
For comparison, at different values of $L$, the fitted distributions divided by $f(\overline{x})$ are shown in Fig. \ref{f:Fig8}. It is also useful to present the mode of the fitted distributions as a function of $L$ in Fig. \ref{f:Fig9}. We trust that the mode converges; there is no reason to believe otherwise.


\begin{figure}[ht]
\includegraphics*[angle=-90,width=8cm]{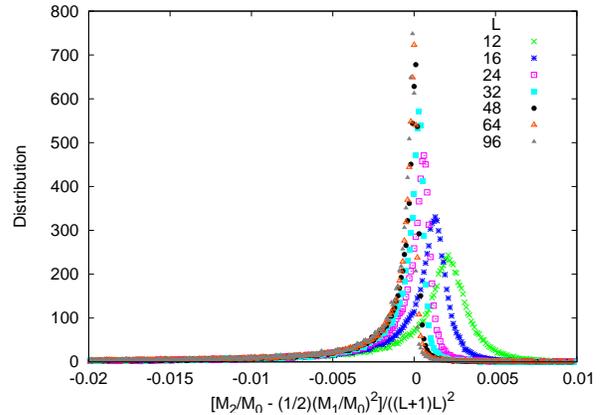}
\caption{\label{f:Fig10}(Color online) Distributions for the second term in the specific heat with various values of lattice size $L$ with $(L+1)\times L$ spins. Each includes $10^5$ disorder realizations, except for $L=96$ which has $30,000$.}
\end{figure}

\begin{figure}[ht]
\includegraphics*[angle=-90,width=8cm]{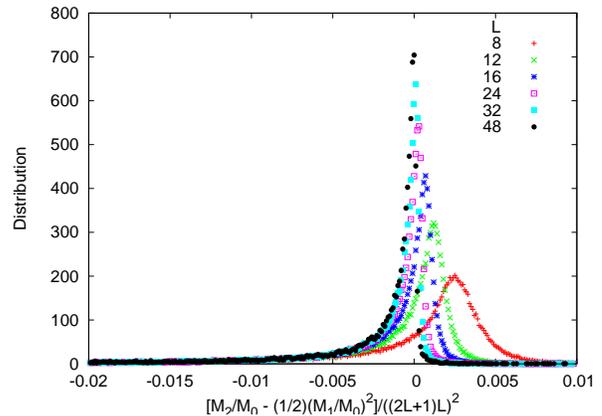}
\caption{\label{f:Fig11}(Color online) Distributions for the second term in the specific heat with various values of lattice size $L$ with $(2L+1)\times L$ spins. Each includes $10^5$ disorder realizations.}
\end{figure}

\begin{figure}[ht]
\includegraphics*[angle=-90,width=8cm]{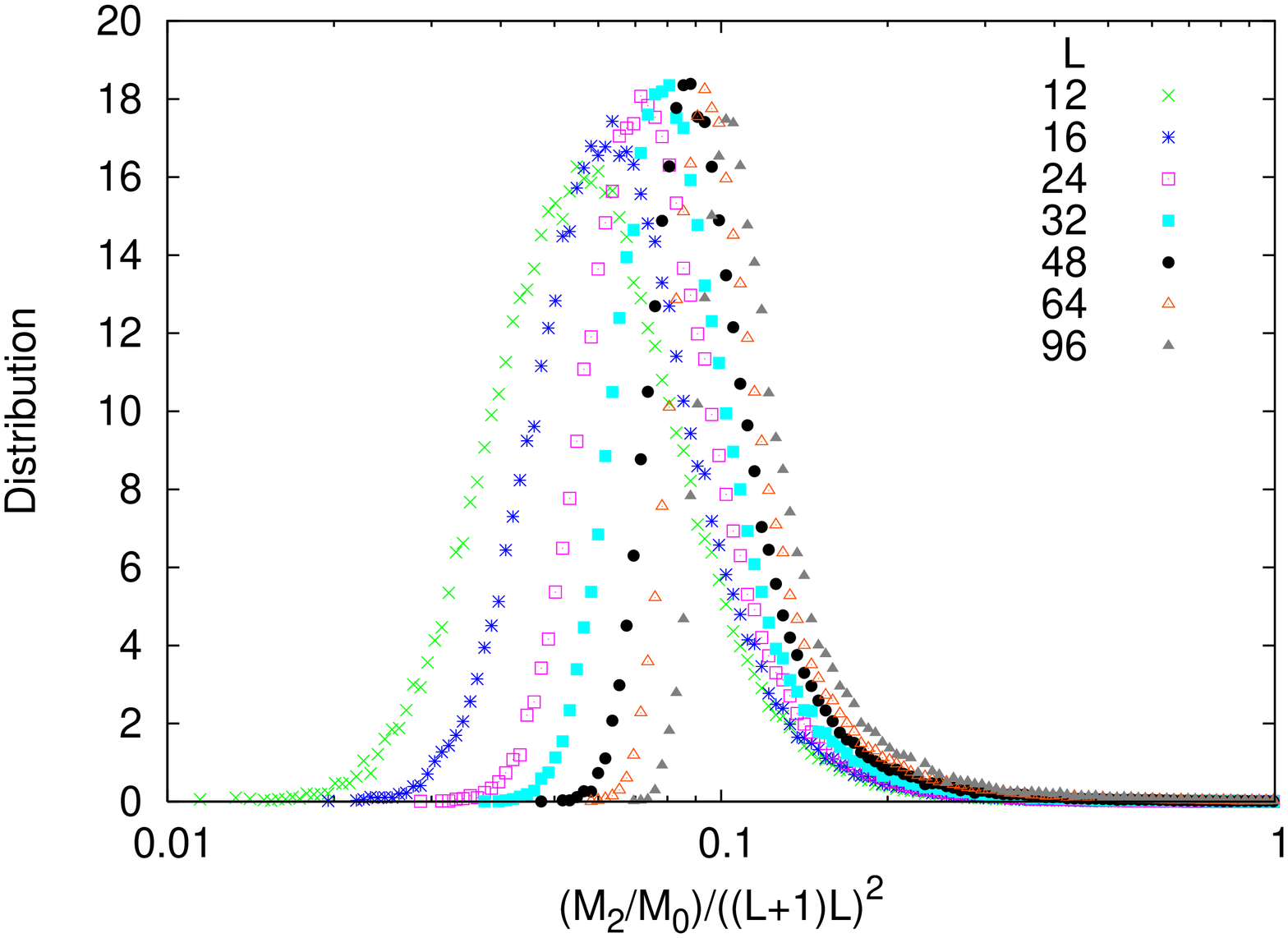}
\caption{\label{f:Fig12}(Color online) Distributions for $\frac{M_2}{M_0}(\frac{1}{(L+1)\times L})^2$ with various values of lattice size $L$ with $(L+1)\times L$ spins. Each distribution includes $10^5$ disorder realizations, except for $L=96$ which has $30,000$.}
\end{figure}

\begin{figure}[ht]
\includegraphics*[angle=-90,width=8cm]{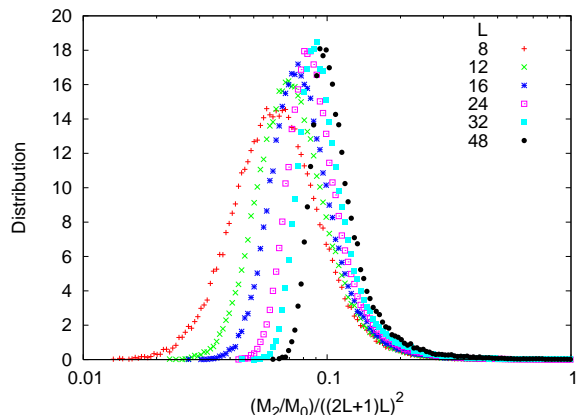}
\caption{\label{f:Fig13}(Color online) Distributions for $\frac{M_2}{M_0}(\frac{1}{(2L+1)\times L})^2$ with various values of lattice size $L$ with $(2L+1)\times L$ spins. Each distribution includes $10^5$ disorder realizations.}
\end{figure}

We have also found distributions for the second contribution to the internal energy, $\frac{M_2}{M_0}-\frac{1}{2}(\frac{M_1}{M_0})^2$. These are shown in Figs. \ref{f:Fig10} and \ref{f:Fig11}. The most likely value is very close to zero. The skewness also suggests the dominance of the first excitations. We thus believe that higher excitations are unlikely to change our conclusion that the energy gap is $2J$. The mode of $\frac{M_2}{M_0}$ alone scales as $((L+1)L)^2$ and $((2L+1)L)^2$ as shown in Figs. \ref{f:Fig12} and \ref{f:Fig13}, respectively.

% START JULIAN
\section{conclusions}
In conclusion, we have reported exact results for the excitations of the
bimodal Ising spin glass on the brickwork lattice by expanding
in arbitrary temperature from the ground state. This is complimentary to
the more usual extrapolation from finite temperature.

We find that the energy gap is $2J$ for both finite and infinite lattices.
The thermodynamic limit is trivial in contrast to the difficulties associated
with the square lattice. Our result may suggest that a $2J$ energy gap is
universal for planar bimodal Ising spin glasses in the thermodynamic limit.
For instance, the triangular lattice could very well be expected to behave
similarly to the square lattice since its plaquettes can be colored using
just two colors; the brickwork lattice requires three colors.

As a final note, we expect a correlation length $\xi \sim \exp(2J/kT)$ in
probable agreement \cite {H01,KL04,SK93,Ratee,Wanyok} with the square lattice.
Our reasoning is based on the construction of correlation functions using
reciprocal defects \cite {GH64,B82,PB05} and closed polygons. The essential
point is that, for a finite lattice, the correlation functions must be
analytical functions of $t$ and thus of $\delta \sim \exp(-2J/kT)$. Comparison
with the asymptotic expression $\exp(-R/\xi)/R^{\eta}$ for
correlation functions at large separation $R$, allows us to deduce that
$\xi^{-1}$ is also an analytical function of $\delta$. If the thermodynamic
limit is trivial for the energy gap, then it is very likely to be so for
the correlation length. It is also known that the fully frustrated brickwork
lattice has a constant correlation length \cite {WZ82} and that the ground
state is not critical, unlike the Villain model \cite {Villain} that
has a nonanalytical free energy although $\xi^{-1} \sim \delta$ nevertheless
\cite{Europhys}.
\section*{ACKNOWLEDGEMENTS}
W. A. thanks the Commission on Higher Education Staff Development Project, Thailand for a scholarship.
Some of the computations were performed on the Tera Cluster at the Thai National Grid Center.


\begin{references}

\bibitem{HY01}
A. K. Hartmann and A. P. Young, Phys. Rev. B {\bf 64}, 180404(R)
(2001).

\bibitem{H01}
J. Houdayer, Eur. Phys. J. B {\bf 22}, 479 (2001).

\bibitem{WS88}
J.-S. Wang and R. H. Swendsen, Phys. Rev. B {\bf 38}, 4840 (1988).

\bibitem{LGM04}
J. Lukic, A. Galluccio, E. Marinari, O. C. Martin, and G. Rinaldi,
Phys. Rev. Lett. {\bf 92}, 117202 (2004).

\bibitem{W05}
J.-S. Wang, Phys. Rev. E {\bf 72}, 036706 (2005).

\bibitem{KLC05}
H. G. Katzgraber, L. W. Lee, and I. A. Campbell, cond-mat/0510668 (2005).

\bibitem{Wanyok}
W. Atisattapong and J. Poulter, New J. Phys. {\bf 10}, 093012 (2008).

\bibitem{SK93}
L. Saul and M. Kardar, Phys. Rev. E {\bf 48}, R3221 (1993);
Nucl. Phys. B {\bf 432}, 641 (1994).

\bibitem{JLM06}
T. J\"{o}rg, J. Lukic, E. Marinari, and O. C. Martin, Phys. Rev. Lett. {\bf 96},
237205 (2006).

\bibitem{KLC07}
H. G. Katzgraber, L. W. Lee, and I. A. Campbell, Phys. Rev. B {\bf 75},
014412 (2007).

\bibitem{VL97}
E. E. Vogel and W. Lebrecht, Z. Phys. B {\bf 102}, 145 (1997).

\bibitem{Anuchit}
A. Aromsawa, Ph.D. Thesis, Mahidol University (2007).

\bibitem{Bendisch}
J. Bendisch, Physica A {\bf 359}, 399 (2006).

\bibitem{deQ06}
S. L. A. de Queiroz, Phys. Rev. B {\bf 73}, 064410 (2006).

\bibitem{NO06}
H. Nishimori and M. Ohzeki, J. Phys. Soc. Jpn. {\bf 75}, 034004 (2006).

\bibitem{Ohzeki}
M. Ohzeki, arXiv:0811.0464v1 (2008).

\bibitem{GH64}
H. S. Green and C. A. Hurst, {\it Order-Disorder Phenomena}
(Interscience, London, 1964).

\bibitem{B82}
J. A. Blackman, Phys. Rev. B {\bf 26}, 4987 (1982).

\bibitem{BP91}
J. A. Blackman and J. Poulter, Phys. Rev. B {\bf 44}, 4374 (1991).

\bibitem{PB05}
J. Poulter and J. A. Blackman, Phys. Rev. B {\bf 72}, 104422
(2005).
\bibitem{Hosking}
J. R. M. Hosking, Appl. Stat. {\bf 34}, 301 (1985). 

\bibitem{LMmethod}
W. H. Press, S. A. Teukolsky, W. T. Vettering, and B. P. Flannery, {\it Numerical Recipes in Fortran}
(Cambridge University Press, Cambridge, 1992).


\bibitem{KL04}
H. G. Katzgraber and L. W. Lee, Phys. Rev. B {\bf 71}, 134404
(2005).

\bibitem{Ratee}
R. Sungthong and J. Poulter, J. Phys. A {\bf 36}, 6347 (2003). 

\bibitem{WZ82}
W. F. Wolff and J. Zittartz, Z. Phys. B {\bf 49}, 139 (1982).

\bibitem{Villain}
J. Villain, J. Phys. C {\bf 10}, 1717 (1977).

\bibitem{Europhys}
J. Lukic, E. Marinari, and O. C. Martin, Europhys. Lett. {\bf 73},
779 (2006).
\end{references}
\end{document}